# Deep Learning Based Detection of Enlarged Perivascular Spaces on Brain MRI

*Tanweer Rashid[a,b,$], Hangfan Liu[a,b,$], Jeffrey B. Ware[c], Karl Li[a], Jose Rafael Romero[d], Elyas Fadaee[a], Ilya M. Nasrallah[b,c], Saima Hilal[e], R. Nick Bryan[c,f], Timothy M. Hughes[g], Christos Davatzikos[c], Lenore Launer[h], Sudha Seshadri[a], Susan R. Heckbert[i], Mohamad Habes[a,b,]\**

[a]*Neuroimage Analytics Laboratory and Biggs Institute Neuroimaging Core, Glenn Biggs Institute for Neurodegenerative Disorders, University of Texas Health Science Center at San Antonio, San Antonio, TX, USA*

[b]*Center for Biomedical Image Computing and Analytics, University of Pennsylvania, Philadelphia, PA, USA*

[c]*Department of Radiation Oncology, University of Pennsylvania, Philadelphia, PA, USA*

[d]*Department of Neurology, School of Medicine, Boston University, Boston, MA, USA*

[e]*Saw Swee Hock School of Public Health, National University of Singapore and National University Health System, Singapore*

[f]*Department of Diagnostic Medicine, Dell Medical School, University of Texas at Austin, Austin, TX, USA*

[g]*Department of Internal Medicine and Department of Epidemiology and Prevention, Wake Forest School of Medicine, Winston-Salem, NC, USA*

[h]*Laboratory of Epidemiology and Population Sciences, National Institute on Aging, National Institutes of Health, Bethesda, MD, USA*

[i]*Department of Epidemiology and Cardiovascular Health Research Unit, University of Washington, Seattle, WA, USA*

*$ Equal contributing first authors.*

*\*Corresponding author. E-mail address: habes@uthscsa.edu*

## ABSTRACT

**BACKGROUND AND PURPOSE:** Deep learning has been demonstrated effective in many neuroimaging applications. However, in many scenarios, the number of imaging sequences capturing information related to small vessel disease lesions is insufficient to support data-driven techniques. Additionally, cohort-based studies may not always have the optimal or essential imaging sequences for accurate lesion detection. Therefore, it is necessary to determine which imaging sequences are crucial for precise detection. This study introduces a novel deep learning framework to detect enlarged perivascular spaces (ePVS) and aims to find the optimal combination of MRI sequences for deep learning-based quantification.

**MATERIALS AND METHODS:** We implemented an effective lightweight U-Net adapted for ePVS detection and comprehensively investigated different combinations of information from SWI, FLAIR, T1-weighted (T1w), and T2-weighted (T2w) MRI sequences. The training data included 21 participants, which were randomly selected from the MESA cohort. Participants had ePVS 683 lesions on average. For T1w, T2w, and FLAIR images, the MESA study collected 3D isotropic MRI scans at six different sites with Siemens scanners. Our training data included participants from all these sites and all the scanner models, and the proposed model was applied to the whole brain instead of selective regions.

**RESULTS:** The experimental results showed that T2w MRI is the most important for accurate ePVS detection, and the incorporation of SWI, FLAIR and T1w MRI in the deep neural network had minor improvements in accuracy and resulted in the highest sensitivity and precision (sensitivity =0.82, precision =0.83). The proposed method achieved comparable accuracy at a minimal time cost compared to manual reading.

**CONCLUSIONS:** The proposed automated pipeline enables robust and time-efficient readings of ePVS from MR scans and demonstrates the importance of T2w MRI for ePVS detection and the potential benefits of using multimodal images. Furthermore, the model provides whole-brain maps of ePVS, enabling a better understanding of their clinical correlates compared to the clinical rating methods within only a couple of brain regions.

## 1. Introduction

Enlargement of perivascular, or Virchow-Robin, spaces [1,2] can manifest cerebral small vessel disease and dysfunction of perivascular drainage routes. Perivascular spaces are fluid-filled spaces that surround arteries, arterioles, veins, and venules [3] in the brain. They are generally microscopic but may become enlarged and visible with increasing age and/or pathologies, i.e., enlarged perivascular spaces (ePVS) [2-5]. Typically, ePVS appear as bright or hyperintense linear or curvilinear structures when running parallel to the imaging plane and ellipsoidal or dot-like when perpendicular to the imaging plane on T2-weighted (T2w) MRI [2,3]. When perivascular spaces are enlarged, they become visible on routine structural MRI, typically with a diameter of less than 3mm. They can reach up to 10-20 mm in regions such as the basal ganglia [3]. While ePVS can be evaluated on T1-weighted (T1w) and T2w sequences, they are easier to visualize and quantify using T2w imaging [6,7].

Many detection/segmentation methods have been proposed [5,7-10,] which rely on T2w exclusively for the detection/segmentation of ePVS. However, it is still unclear if models depending on a single modality such as T2w could account for similar-appearing brain lesions such as white matter hyperintensities (WMH), lacunes, and infarcts. WMH are hyperintense on T2w sequences and can appear as isointense or hypointense on T1w sequences; lacunes are round or ovoid subcortical fluid-filled cavities of between 3 mm and 15 mm in diameter, while infarcts are neuroimaging evidence of recent infarction in the territory of one perforating arteriole [3].

In this paper, we aim to evaluate the feasibility and effectiveness of an automated deep learning-based method for segmenting ePVS using multiple MRI sequences from a subset of participants in the Multi-Ethnic Study of Atherosclerosis (MESA) cohort. The brain data collected by the MESA Atrial Fibrillation (AFib) [11-14] ancillary study at Exam 6 offer a unique and rich dataset of high-quality brain MRI at clinical field strength and high spatial resolution (1 mm isotropic images). We aim to evaluate the accuracy and reliability of ePVS segmentation in the presence or absence of T2w MRI and when T2w is combined with other MRI sequences. We used a variation of our method, previously developed using MESA brain MRI data for fully automated detection of cerebral microbleeds and non-hemorrhage iron deposits in the basal ganglia [15], and investigate the optimal strategy of combining information from SWI, FLAIR, T1w, and T2w MRI sequences. A set of ePVS segmentations by a human expert served as the gold standard for model training.

Automation is ideal in large cohort studies for feasibility, improved reproducibility, and reduced human error [16]. Accurate and reliable methods are also essential for deriving rich datasets from large cohorts to study associations with demographic, cognitive, and vascular risk factors [17-19], or to refine the development of new methods [20,21].

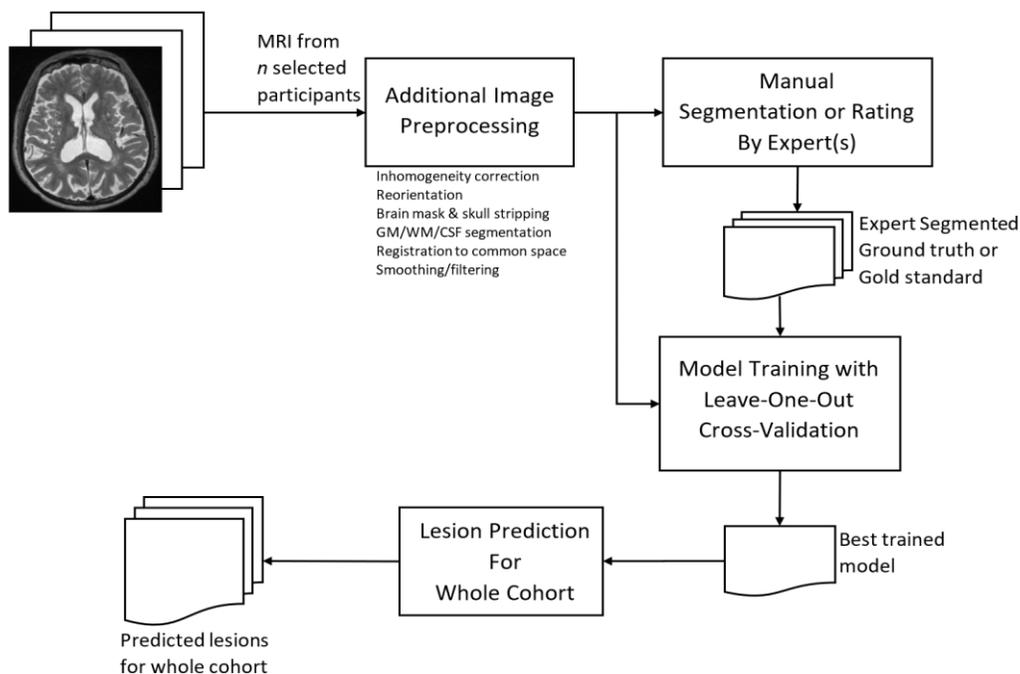

**Fig. 1.** Overview of the ePVS detection/segmentation procedure.

To our knowledge, this study is among the first to comprehensively evaluate multimodal imaging for ePVS detection with deep learning. The main contributions of this paper include the following:

1. Development of an effective deep learning scheme with data fusion for accurate ePVS segmentation.

2. Application of the proposed model to the whole brain instead of selective regions.

3. Investigation of the use of different sequences for optimal performance.

## 2. Related Works

Previous ePVS segmentation methods typically adopt conventional machine learning techniques such as vessel enhancement filters [7] and support vector machines (SVM) [22]. Ballerini et al. trained a model on T2-contrast MR images [7] and evaluated it by categorical scores [6]. González-Castro et al. applied an SVM classifier with a bag of visual words-based descriptors to the T2-weighted MR images with a focus on the basal ganglia [22]. Wang et al. developed a semi-automatic computational method that extracts ePVS on bilateral ovoid basal ganglia on intensity-normalized T2w MRI [9]. Meanwhile, some works exploited handcrafted features as predictors. For example, Boespflug et al. used signal intensities and morphologic characterizations, including width, volume, and linearity [23], while Ramirez et al. used set localized intensity thresholds for quantification of perivascular spaces [24], and Zhang et al. proposed vascular feature based structured learning for 3-dimensional ePVS segmentation using T2w data [8]. Besides, to facilitate these models, Sepehrband et al. combined T1- and T2w images to enhance PVS contrast and intensify visibility [25].

With the recent success of deep learning techniques [21,26-28], some deep neural network models were proposed for ePVS segmentation. For instance, Boutinaud et al. developed a deep learning algorithm based on an autoencoder and a U-shaped network for the 3-dimensional segmentation of ePVS in deep white matter and basal ganglia using T1-weighted MRI data [29]. Lian et al. proposed a fully convolutional neural network using 7T T2-weighted MRI for efficient segmentation of ePVS [10], Dubost et al. implemented separate convolutional neural networks for midbrain, hippocampi, basal ganglia, and centrum semiovale, trained on T2-contrast MRI to quantify PVS [30], Sudre et al. redesigned the region-based convolutional neural networks model to jointly detect and characterize markers of age-related neurovascular changes [31]. Jung et al. presented a deep 3-dimensional convolutional neural network with densely connected networks with skip connections for ePVS enhancement of 7T MRI [32]. In general, these works did not investigate how to fully utilize different sources of information for improved ePVS detection on the whole brain based on deep data-driven techniques. Furthermore, these prior studies mostly used 7T MRI, which is highly experimental and not readily available. 3T MRI is more conventional and widely available, so a deep learning model tailored for 3T MRI a more cost-effective choice.

## 3. Materials and Methods

The key point of the proposed scheme is a deep fusion of information from different MRI sequences. An overview of the whole procedure is summarized in Fig. 1. Standard image processing techniques were first applied to raw MRI data of different sequences from a subset of the MESA cohort, including inhomogeneity correction, reorientation, smoothing and filtering, brain masking and skull-stripping, followed by gray matter (GM), white matter (WM) and cerebrospinal fluid (CSF) segmentation. Then the participants' MRIs were registered to SWI. Preprocessed MRI data were manually segmented to obtain the ground truth used for model training with leave-one-out cross-validation.

### 3.1. Data

The training data included 21 participants, which were randomly selected from the MESA cohort. For T1w, T2w, and FLAIR images, the MESA study collected 3D isotropic MRI scans at six different sites with Siemens scanners (Skyra with a 20-channel head coil, Prisma and Prisma Fit with a 32-channel head coil). Our training data included participants from all these sites and all the scanner models, thus ensuring generalizability within the MESA cohort. The MRI scan parameters are shown in supplementary material Table S1.

The ages of the 21 participants range from 64 to 94 years, with an average of 78.7 years, and 12 of them are female. The average total number of individual lesions per participant is more than 683. The ePVS segmentation of these participants was performed by an experienced radiologist (JBW) and served as ground truth for model training (see supplementary material Table S2). The average time to create the ground truth was ~24 hours per scan. The manual segmentation was performed using co-registered T2w, T1w, and FLAIR images to ensure a reduced likelihood of false positives such as WMH or lacunes [3] being present in the ground truth. The average time needed to complete a manual segmentation was around 30 hours per participant. For model training and evaluation, we used FLAIR, SWI, T1w, and T2w images, which were reoriented, N4 bias corrected [33,] and skull-stripped [34]. The SWI phase mask was generated from the phase images using a high-pass filter of size 64 x 64 to remove artifacts, and the SWI was generated by multiplying the magnitude image with the phase mask[35]. For the creation of the reference annotation and machine-based inference, only the SWI image with the shortest echo time (TE=7.5 ms) was used since SWI acquired with longer echo times are noisier. Examples of ePVS on the different sequences are shown in Fig. 2. The MRI scans used in this study have a high spatial resolution, making it possible to detect small ePVS. However, in clinical settings, the slice thickness is larger to allow for less scan time, so small lesions occurring between slices may not be visible.

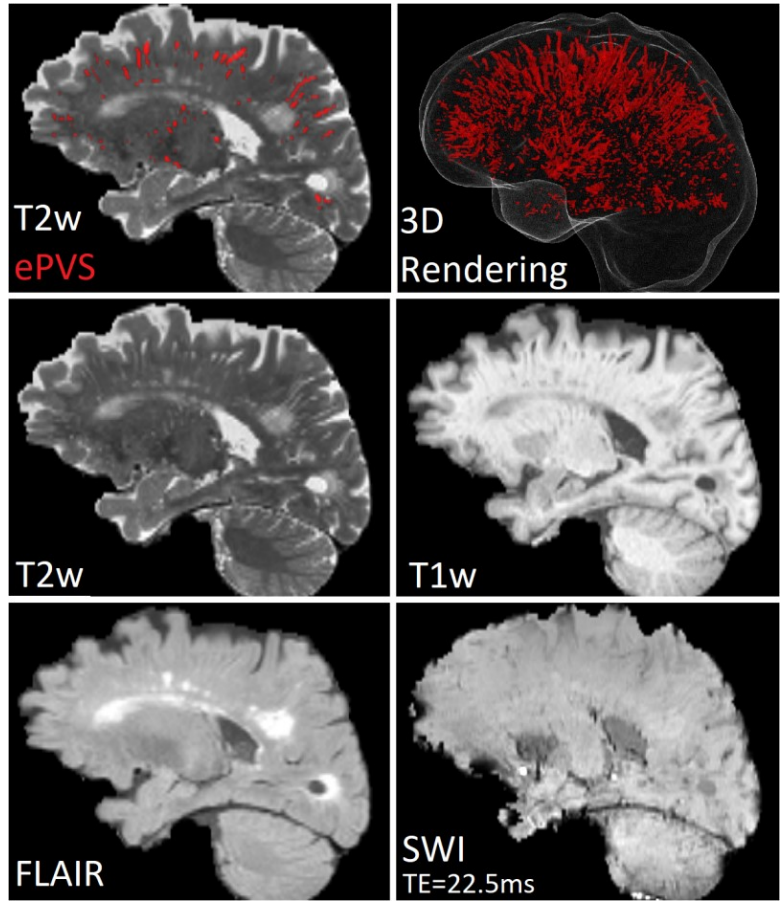

**Fig. 2.** Examples of ePVS in different MRI sequences.

### 3.2. Deep Fusion of Different Sequences

Suppose $f_U: \mathbb{R}^{n \times S} \to \mathbb{R}^S$ is a nonlinear function with a set of learnable parameters $U$, where $n$ is the number of MRI Sequences used, and $S$ is the size of the images, $f$ maps the $n$ images to voxel-wise labels indicating whether the voxel contains ePVS or not. In this study, $f_U$ is implemented as a multi-channel deep neural network [15], a variation of the standard U-Net [36] and has been demonstrated superior to conventional U-Net for small lesions [15]. A typical U-Net comprises a down-sampling or encoding path and a symmetric up-sampling or decoding path. The down-sampling course consists of a series of convolutional blocks, normalization blocks, activation blocks, and pooling blocks. The up-sampling path consists of a series of convolutional blocks, normalization blocks, activation blocks, and transposes convolutional blocks. The feature maps of each corresponding down-sampling path and up-sampling path are concatenated.

The proposed scheme could perform a deep fusion of information from different sequences. The ePVS detection/segmentation model fuses information from T2w, SWI, FLAIR, and T1w images through the multi-channel U-Net. It was designed in a scalable manner, i.e., the network using T2 only was basically a single-channel U-Net, and can be easily expanded to include multiple sequences. The manual segmentations by the human expert were used to train the deep learning model using leave-one-out cross-validation. Specifically, in each iteration of the leave-one-out cross-validation, we use data from 20 subjects for network training and data from one subject for testing. Also, from the 20 subjects used in training, four were used exclusively for within-training validation.

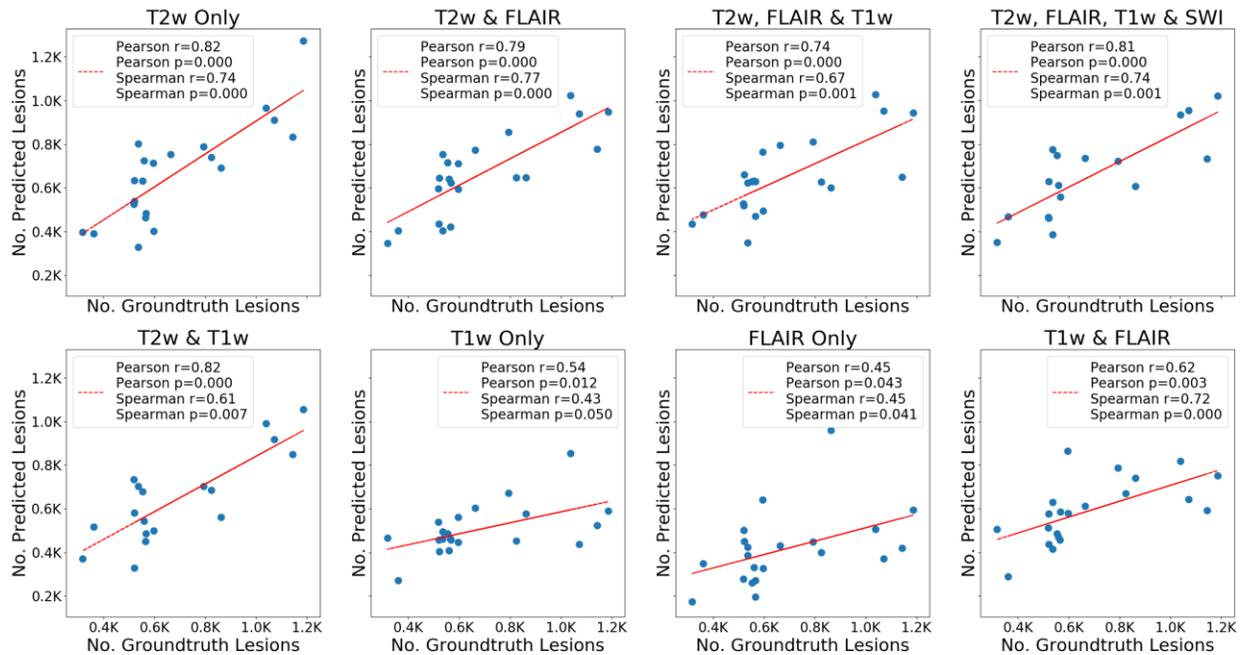

**Fig. 3.** Scatterplots of number of lesions/per case.

We aimed to train multiclass models, where predicted classes were background and ePVS, using the following combinations of imaging sequences: (1) T2w-only, (2) T2w and FLAIR, (3) T2w, T1w and FLAIR, (4) T2w, T1w, FLAIR, and SWI, (5) T2w and T1w, (6) FLAIR only, (7) T1w only, (8) T1w and FLAIR. Each 3-dimensional (3D) scan was cut into 2D axial slices, which underwent data augmentation through geometric operations such as flipping, translation, and rotation. For example, a single T2w MRI image having 96 axial slices resulted in 23880 axial slices after data augmentation. These augmented data were fed into the neural network as data samples.

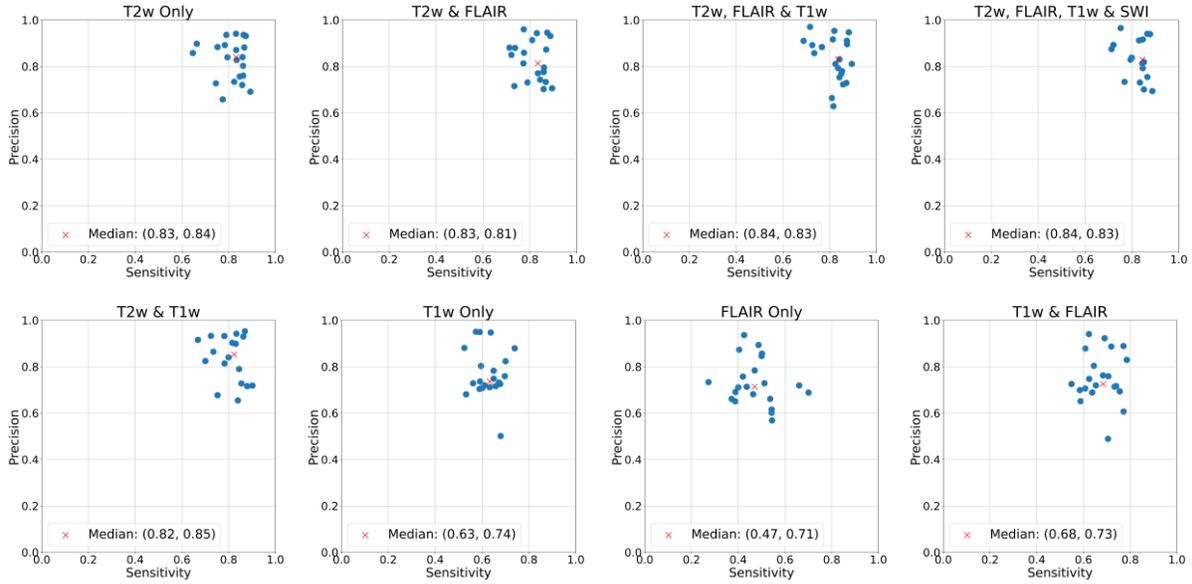

**Fig. 4.** Sensitivity vs precision.

### 3.3. Analysis of Detection Results

The accuracy of these models was based on three parameters: sensitivity $S$, precision $P$, and magnitude accuracy $A$, which are defined as

$$S = TP/(TP + FN), \ P = TP/(TP + FP), \ A = \sqrt{S^2 + P^2},$$

where TP is the number of true positives, FN is the number of false negatives, and FP stands for false positives.

We also selected metrics effective for small lesions like ePVS, where shape information and volume are essential. The ePVS could be as small as one voxel. The analysis included Bland-Altman plots and scatterplots of ePVS count and volumes (prediction vs. expert labeled data), as well as sensitivity and precision based on the center of mass of the lesions. We also assessed performance using intra-class correlation coefficients (ICC) [37], volumetric similarity [38], area under the curve (AUC) from receiver operator curves, Hausdorff distance [39,] and Mahalanobis distance [40]. For ICC, we calculate the correlation between the total number of lesions of ground truth and that of the prediction, as well as the correlation between the total volumes of ground truth and that of the prediction. Hausdorff distance calculates the distance between two point sets that correspond to ground truth labels and segmentations, respectively, while Mahalanobis distance is a multivariate distance metric that measures the distance between a point and a distribution and is particularly effective for classification on highly imbalanced datasets. The mean metrics are obtained by averaging over subjects, e.g., suppose $S_i$ is the sensitivity obtained by testing subject $i$ ($i = 1, 2, \ldots, 21$), then average sensitivity $\bar{S} = \frac{1}{21} \sum_{i=1}^{21} S_i$.

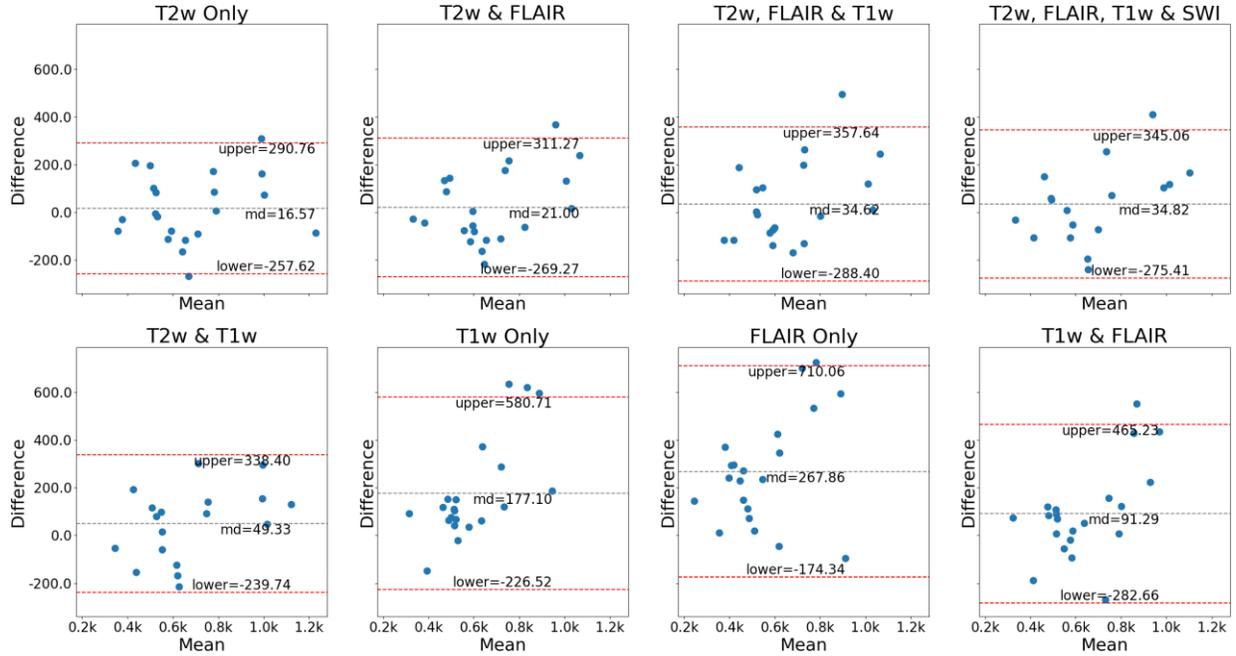

**Fig. 5.** Bland-Altman plot of number of lesions.

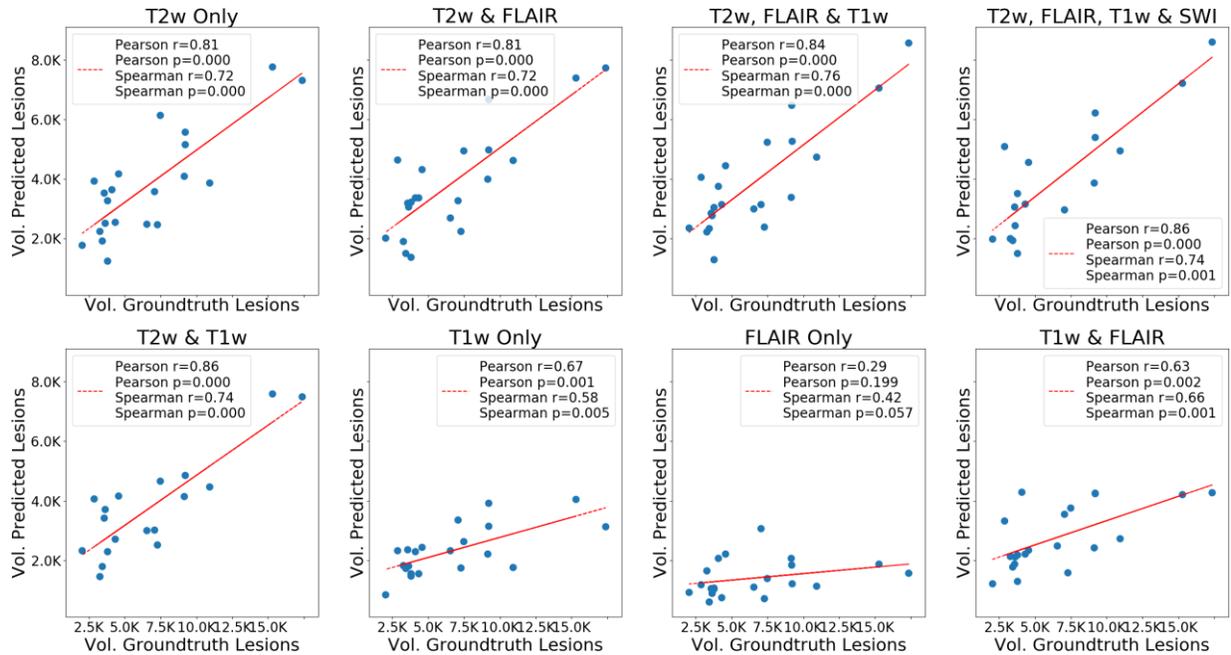

**Fig. 6.** Scatterplots of volume of lesions.

## 4. Results

The mean evaluation metrics with corresponding standard errors of all subjects, including sensitivity, precision, magnitude accuracy, ICC, volumetric similarity, AUC, Hausdorff distance, and Mahanabolis distance, are shown in Table 1. The results indicate that T2w MRI is the most informative, with the best performance of any single sequence and near-optimal for several measures. For most measures, the combination of T2w, FLAIR, T1w, and SWI achieved the best performance. Adding SWI to the combination of the other three sequences offered minimal overall gain but improved ICC.

Fig. 3 displays the correlations between the number of predicted lesions and that of ground-truth lesions. The highest correlations are achieved by using T2w. Fig. 4 plots the points located by pairs $(S, P)$ from all the participants and indicates that by including T2w, FLAIR, and T1w the model could attain the highest magnitude accuracy, which is reflected by the distance between $(\bar{S}, \bar{P})$ and $(0, 0)$ in the figure, where $\bar{S}$ and $\bar{P}$ are the median sensitivity and median precision, respectively.

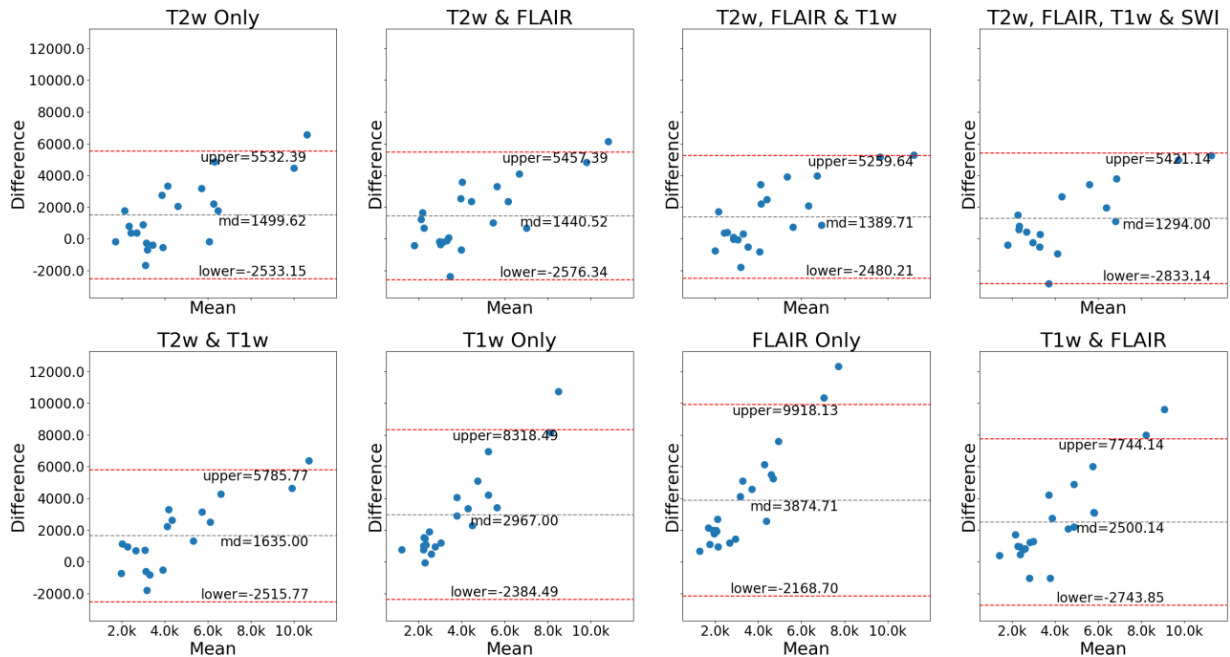

**Fig. 7.** Bland-Altman plot of volume of lesions.

Fig. 5 shows the Bland-Altman plots of number of lesions, demonstrating that the mean difference between the prediction and the gold standard as well as the random fluctuations around the mean, reached the minimum when using T2w only and remained low when incorporating other sequences. Fig. 6 displays the correlations between the volume of predicted lesions and the volume of the ground truth, reaffirming the importance of using T2w for ePVS segmentation. Fig. 7 shows the Bland-Altman plots of lesion volumes, indicating that the combination of T2w, FLAIR, T1w, and SWI could attain better results than using FLAIR, T1w only since the mean difference and the fluctuations were minimal when combining T2w, T1w, FLAIR, and SWI, and were significantly smaller when T2w is included.

**Table 1 -** Subject-wise evaluation. The best scores are marked as red and the second best as blue.

| Expts | Avg Sensitivity | Avg Precision | Avg Mag Accuracy | Avg Volumetric Similarity | Avg AUC | Average Hausdorff Distance | Average Mahanabolis Distance | ICC (#Lesions) | ICC (Volume |
|---|---|---|---|---|---|---|---|---|---|
| T2w | 0.81±0.01 | 0.83±0.02 | 1.16±0.02 | 0.81±0.03 | 0.72±0.01 | 1.41±0.08 | 0.17±0.02 | 0.83 | 0.59 |
| 2w+FLAIR | 0.82±0.01 | 0.82±0.02 | 1.16±0.01 | 0.81±0.03 | 0.73±0.01 | 1.38±0.08 | 0.16±0.02 | 0.77 | 0.60 |
| 2w+FLAIR +T1w | 0.82±0.01 | 0.83±0.02 | 1.17±0.01 | 0.82±0.03 | 0.74±0.01 | 1.27±0.07 | 0.17±0.02 | 0.70 | 0.63 |
| 2w+FLAIR +T1w+SWI | 0.82±0.02 | 0.83±0.02 | 1.17±0.02 | 0.82±0.03 | 0.74±0.01 | 1.28±0.07 | 0.16±0.01 | 0.77 | 0.67 |
| T2w+T1w | 0.80±0.02 | 0.84±0.02 | 1.16±0.02 | 0.78±0.02 | 0.71±0.01 | 1.40±0.09 | 0.18±0.01 | 0.79 | 0.58 |
| T1w | 0.63±0.01 | 0.77±0.02 | 1.00±0.02 | 0.66±0.04 | 0.59±0.01 | 2.49±0.11 | 0.24±0.02 | 0.30 | 0.18 |
| FLAIR | 0.47±0.02 | 0.73±0.02 | 0.88±0.02 | 0.48±0.04 | 0.53±0.02 | 3.59±0.09 | 0.35±0.04 | 0.24 | 0.05 |
| 1w+FLAIR | 0.68±0.02 | 0.75±0.02 | 1.02±0.02 | 0.71±0.03 | 0.61±0.01 | 2.30±0.11 | 0.23±0.02 | 0.50 | 0.25 |

Based on such observations, we can see that although T1 and FLAIR are more standard research sequences, ePVS ratings using these two sequences are not nearly as accurate as including T2w, and incorporating other sequences did not improve results significantly. However, using information from different modalities enables the model to effectively distinguish ePVS from mimics like white matter lesions and lacunes, as demonstrated in Fig. 8.

# 5. Discussion and Conclusions

Enlarged perivascular spaces (ePVS) are increasingly recognized as a subclinical biomarker for brain health and disease, including cerebrovascular disease, and therefore quantification is of interest to the research community. Manual quantification of individual ePVS is extremely time-consuming [9,24], operator-dependent, and may not accurately reflect the actual burden of ePVS. Data-driven automated systems, including deep learning models, provide a promising way to generate robust, reproducible, and rapid quantification of ePVS from brain MRI scans.

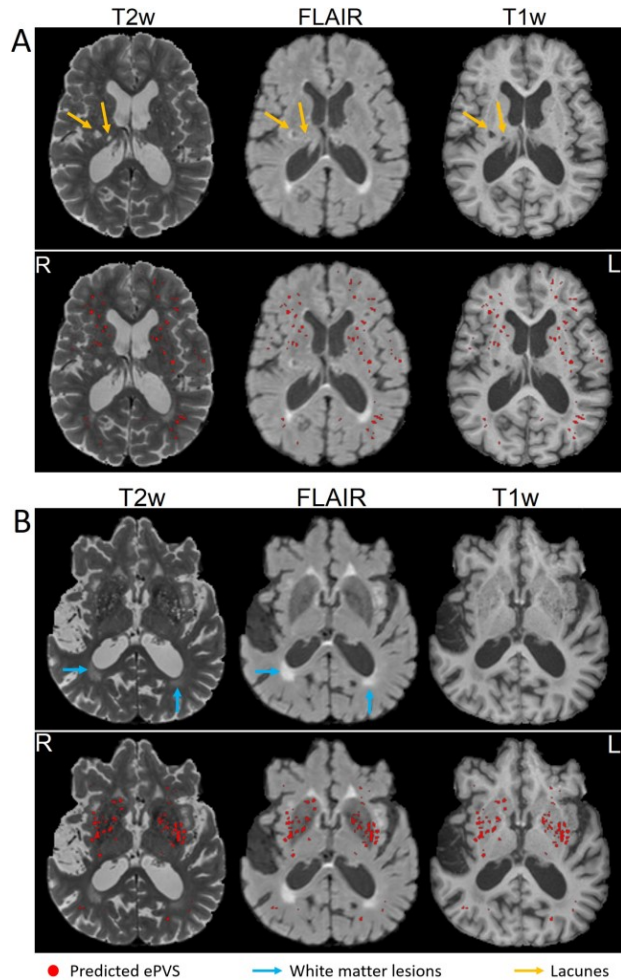

**Fig. 8.** An example of the predicted ePVS.

Automated ePVS quantification is challenging due to the existence of mimics like lacunes and white matter lesions, which may lead to false positive measurements. Furthermore, in many scenarios, the number of neuroimaging data samples could be insufficient to support data-driven systems. Such problems still remain in recently published deep learning methods.

In general, there are several limitations: 1) The issue of insufficient data samples is not addressed, and consequently there could be problems like overfitting; 2) It is still under question as to whether one single modality could be sufficiently informative for ePVS quantification; 3) The advantages of combining different sequences for the application is not investigated; 4)

Existing methods generally use 7T MRI while 3T data is more common in practice; 5) Existing methods were only applied to particular regions rather than the whole brain.

To address these issues, this study aims to fully exploit the informative 3T MRI data available by jointly utilizing different sequences and investigating the optimal strategy of fusing information from different sequences in the deep learning framework for ePVS segmentation, which could be applied to the whole brain. Specifically, since the number of data samples is often limited, it is of great importance to make full use of the data available, and the fusion of information from different sequences could be an effective solution. The deep learning model adopts a lightweight multi-channel variation of the U-Net tailored for the application. The experimental results demonstrate that the combination of T2w, FLAIR, T1w, and SWI leads to the best segmentation performance and that performance with T1w alone is worse than T2w alone for detecting ePVS. Our results suggest that if the quantification of ePVS is of interest, prospective research studies should include T2w imaging in a brain MRI protocol. T1w images, which are by far more prevalent in research studies due to their utility in brain tissue segmentation, should be expected to provide less accurate quantification of ePVS.

For regional evaluation, we derived several regions based on the existing MUSE segmentations, as shown in Fig. S1 of the supplementary material. Based on these regions, we did the same sensitivity and precision calculations for each individual region in all the experiments. The metrics, including mean sensitivity and precision, etc., are in Table S3 ~ S9 of the supplementary material. We can see that in the basal ganglia, the sensitivity and precision are high for all experiments, even when using only T1 or FLAIR. This suggests we can get reliable and accurate ePVS readings in the basal ganglia using only T1 and/or FLAIR. On the other hand, we see that the sensitivity and precision are poor in the hippocampus and temporal regions. This is because of false positives due to the presence of blood vessels prevalent in those regions. Currently, the most clinically relevant regions for ePVS readings are the basal ganglia, centrum semiovale, and maybe the midbrain [2,3]. So our experiments show that our models can make accurate predictions in the basal ganglia and the centrum semiovale, even when T2 is absent.

In conclusion, the proposed automated pipeline enables robust and time-efficient readings of ePVS from MR scans and demonstrates the importance of T2w MRI for ePVS detection and the insignificant benefit of using multimodal images. It may also provide a potential way to alleviate the issues brought about by the limitation of data samples. The automated pipeline will help in generating a rich variable set in MESA that will enable the examination of ePVS in relation to other risk factors. A limitation of the study is that manual ePVS segmentation from only one expert is available.

# Deep Learning Based Detection of Enlarged Perivascular Spaces on Brain MRI

## Supplementary Material

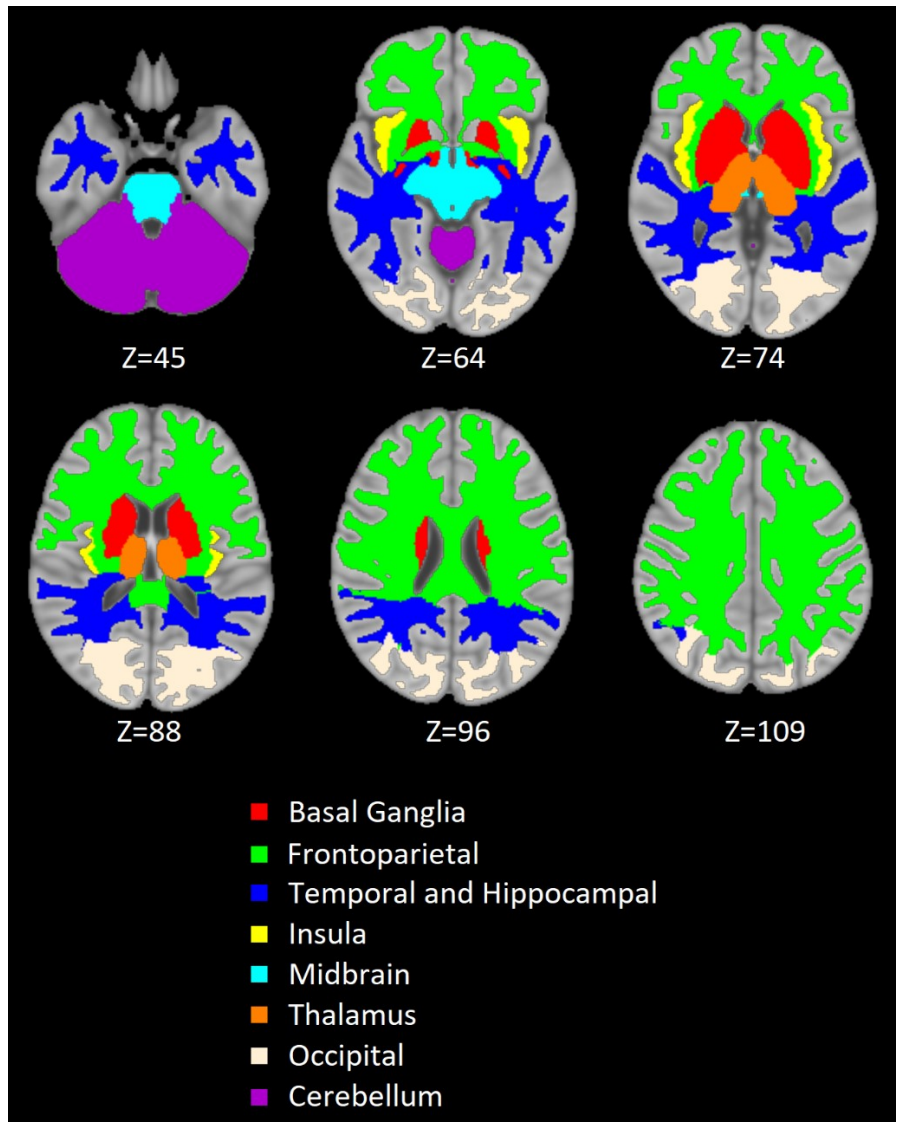

Fig. S1. Regions based on the existing MUSE segmentations.

**Table S1 –** MRI Scanner parameters.

| MRI Modalities | TR (ms) | TE (ms) | FOV (mm) | Flip Angle | Slice Thickness (mm) | No of Slices | Matrix | Scan Duration |
|---|---|---|---|---|---|---|---|---|
| T1w | 1900 | 2.93 | 250 | 9 | 1 | 176 | 256x256 | 4:26 |
| T2w | 3200 | 408 | 250 | 120 | 1 | 176 | 256x256 | 4:08 |
| FLAIR | 6000 (TI=2200) | 289 | 250 | 120 | 1 | 160 | 258x221 | 4:14 |
| SWI | 35 | 7.5, 15, 22.5, and 30 | 256 | 15 | 1.5 | 96 | 256x192 | 6:00 |

**Table S2 –** Number and size of ePVS based on the expert labelled ground truth.

| ID | EPVS Count | Total Voxels | Avg Voxels |
|---|---|---|---|
| Participant 1 | 554 | 2840 | 5.13 |
| Participant 2 | 522 | 3641 | 6.98 |
| Participant 3 | 318 | 2741 | 8.62 |
| Participant 4 | 825 | 5982 | 7.25 |
| Participant 5 | 536 | 2286 | 4.26 |
| Participant 6 | 1072 | 8735 | 8.15 |
| Participant 7 | 794 | 7357 | 9.27 |
| Participant 8 | 1144 | 7303 | 6.38 |
| Participant 9 | 863 | 5643 | 6.54 |
| Participant 10 | 560 | 3020 | 5.39 |
| Participant 11 | 596 | 3261 | 5.47 |
| Participant 12 | 1187 | 13880 | 11.69 |
| Participant 13 | 597 | 5232 | 8.76 |
| Participant 14 | 521 | 2608 | 5.01 |
| Participant 15 | 566 | 5819 | 10.28 |
| Participant 16 | 664 | 7354 | 11.08 |
| Participant 17 | 1039 | 12219 | 11.76 |
| Participant 18 | 567 | 3448 | 6.08 |
| Participant 19 | 519 | 2893 | 5.57 |
| Participant 20 | 536 | 3019 | 5.63 |
| Participant 21 | 361 | 1616 | 4.48 |

**Table S3 -** Subject-wise evaluation in the basal ganglia region. The best scores are marked as red and the second best as blue.

| Expts | Avg Sensitivity | Avg Precision | Avg Mag Accuracy | Avg Volumetric Similarity | Avg AUC | Average Hausdorff Distance | Average Mahanabolis Distance | ICC (#ePVS) | ICC (Volume) |
|---|---|---|---|---|---|---|---|---|---|
| T2w | 0.92 | 0.89 | 1.28 | 0.93 | 0.82 | 0.58 | 0.17 | 0.76 | 0.93 |
| T2w+FLAIR | 0.92 | 0.89 | 1.29 | 0.92 | 0.82 | 0.58 | 0.16 | 0.76 | 0.93 |
| T2w+FLAIR +T1w | 0.92 | 0.89 | 1.28 | 0.92 | 0.82 | 0.56 | 0.19 | 0.63 | 0.96 |
| T2w+FLAIR +T1w+SWI | 0.94 | 0.90 | 1.30 | 0.93 | 0.82 | 0.53 | 0.15 | 0.67 | 0.96 |
| T2w+T1w | 0.88 | 0.88 | 1.28 | 0.93 | 0.80 | 0.61 | 0.19 | 0.63 | 0.92 |
| T1w | 0.84 | 0.83 | 1.19 | 0.86 | 0.66 | 1.33 | 0.30 | 0.55 | 0.67 |
| FLAIR | 0.88 | 0.81 | 1.20 | 0.85 | 0.68 | 1.26 | 0.30 | 0.23 | 0.21 |
| T1w+FLAIR | 0.80 | 0.80 | 1.13 | 0.68 | 0.56 | 2.01 | 0.45 | 0.32 | 0.63 |

**Table S4 -** Subject-wise evaluation in the frontoparietal region. The best scores are marked as red and the second best as blue.

| Expts | Avg Sensitivity | Avg Precision | Avg Mag Accuracy | Avg Volumetric Similarity | Avg AUC | Average Hausdorff Distance | Average Mahanabolis Distance | ICC (#ePVS) | ICC (Volume) |
|---|---|---|---|---|---|---|---|---|---|
| T2w | 0.82 | 0.85 | 1.18 | 0.78 | 0.71 | 1.40 | 0.23 | 0.78 | 0.58 |
| T2w+FLAIR | 0.82 | 0.85 | 1.19 | 0.77 | 0.72 | 1.30 | 0.17 | 0.74 | 0.60 |
| T2w+FLAIR +T1w | 0.83 | 0.85 | 1.19 | 0.80 | 0.74 | 1.24 | 0.19 | 0.67 | 0.64 |
| T2w+FLAIR +T1w+SWI | 0.83 | 0.85 | 1.19 | 0.77 | 0.73 | 1.25 | 0.19 | 0.70 | 0.65 |
| T2w+T1w | 0.77 | 0.85 | 1.17 | 0.75 | 0.71 | 1.43 | 0.21 | 0.61 | 0.61 |
| T1w | 0.64 | 0.79 | 1.02 | 0.67 | 0.58 | 2.57 | 0.34 | 0.24 | 0.18 |
| FLAIR | 0.49 | 0.77 | 0.92 | 0.46 | 0.52 | 3.77 | 0.52 | 0.22 | 0.04 |
| T1w+FLAIR | 0.69 | 0.78 | 1.04 | 0.71 | 0.60 | 2.39 | 0.29 | 0.46 | 0.26 |

**Table S5 -** Subject-wise evaluation in the temporal and hippocampal region. The best scores are marked as red and the second best as blue.

| Expts | Avg Sensitivity | Avg Precision | Avg Mag Accuracy | Avg Volumetric Similarity | Avg AUC | Average Hausdorff Distance | Average Mahanabolis Distance | ICC (#ePVS) | ICC (Volume) |
|---|---|---|---|---|---|---|---|---|---|
| T2w | 0.78 | 0.78 | 1.11 | 0.79 | 0.72 | 1.78 | 0.25 | 0.87 | 0.75 |
| T2w+FLAIR | 0.78 | 0.77 | 1.11 | 0.77 | 0.72 | 1.83 | 0.23 | 0.83 | 0.73 |
| T2w+FLAIR +T1w | 0.78 | 0.78 | 1.11 | 0.80 | 0.74 | 1.62 | 0.24 | 0.80 | 0.75 |
| T2w+FLAIR +T1w+SWI | 0.76 | 0.78 | 1.10 | 0.77 | 0.73 | 1.77 | 0.26 | 0.78 | 0.69 |
| T2w+T1w | 0.73 | 0.80 | 1.11 | 0.77 | 0.70 | 1.86 | 0.27 | 0.84 | 0.72 |
| T1w | 0.54 | 0.68 | 0.88 | 0.65 | 0.58 | 3.45 | 0.39 | 0.40 | 0.30 |
| FLAIR | 0.33 | 0.63 | 0.72 | 0.40 | 0.52 | 5.02 | 0.53 | 0.18 | 0.07 |
| T1w+FLAIR | 0.59 | 0.67 | 0.90 | 0.70 | 0.60 | 3.19 | 0.37 | 0.44 | 0.40 |

**Table S6 -** Subject-wise evaluation in the insula region. The best scores are marked as red and the second best as blue.

| Expts | Avg Sensitivity | Avg Precision | Avg Mag Accuracy | Avg Volumetric Similarity | Avg AUC | Average Hausdorff Distance | Average Mahanabolis Distance | ICC (#ePVS) | ICC (Volume) |
|---|---|---|---|---|---|---|---|---|---|
| T2w | 0.47 | 0.63 | 0.86 | 0.55 | 0.62 | 4.62 | 1.59 | 0.80 | 0.43 |
| T2w+FLAIR | 0.55 | 0.68 | 0.95 | 0.52 | 0.60 | 7.65 | 1.37 | 0.80 | 0.40 |
| T2w+FLAIR +T1w | 0.60 | 0.71 | 0.98 | 0.58 | 0.64 | 5.02 | 1.29 | 0.77 | 0.25 |
| T2w+FLAIR +T1w+SWI | 0.63 | 0.69 | 1.04 | 0.58 | 0.64 | 3.55 | 1.06 | 0.78 | 0.28 |
| T2w+T1w | 0.47 | 0.70 | 0.92 | 0.55 | 0.60 | 9.00 | 1.91 | 0.84 | 0.22 |
| T1w | 0.39 | 0.53 | 0.72 | 0.54 | 0.56 | 6.50 | 3.20 | 0.38 | 0.06 |
| FLAIR | 0.34 | 0.53 | 0.70 | 0.41 | 0.51 | 9.78 | 14.35 | 0.03 | 0.01 |
| T1w+FLAIR | 0.38 | 0.49 | 0.68 | 0.48 | 0.55 | 7.06 | 3.36 | 0.07 | -0.04 |

**Table S7** - Subject-wise evaluation in the midbrain region. The best scores are marked as red and the second best as blue.

| Expts | Avg Sensitivity | Avg Precision | Avg Mag Accuracy | Avg Volumetric Similarity | Avg AUC | Average Hausdorff Distance | Average Mahanabolis Distance | ICC (#ePVS) | ICC (Volume) |
|---|---|---|---|---|---|---|---|---|---|
| T2w | 0.72 | 0.73 | 1.05 | 0.73 | 0.73 | 2.18 | 0.72 | 0.79 | 0.69 |
| T2w+FLAIR | 0.77 | 0.71 | 1.07 | 0.78 | 0.74 | 2.39 | 0.65 | 0.63 | 0.61 |
| T2w+FLAIR +T1w | 0.78 | 0.71 | 1.08 | 0.79 | 0.77 | 2.43 | 0.79 | 0.65 | 0.54 |
| T2w+FLAIR +T1w+SWI | 0.79 | 0.73 | 1.09 | 0.80 | 0.76 | 2.54 | 0.74 | 0.60 | 0.72 |
| T2w+T1w | 0.76 | 0.70 | 1.07 | 0.80 | 0.76 | 2.32 | 0.71 | 0.55 | 0.55 |
| T1w | 0.53 | 0.69 | 0.89 | 0.62 | 0.62 | 3.76 | 1.58 | 0.39 | 0.56 |
| FLAIR | 0.41 | 0.62 | 0.77 | 0.42 | 0.53 | 5.98 | 1.68 | 0.29 | 0.06 |
| T1w+FLAIR | 0.56 | 0.66 | 0.91 | 0.66 | 0.63 | 3.62 | 1.32 | 0.38 | 0.51 |

**Table S8** - Subject-wise evaluation in the thalamus region. The best scores are marked as red and the second best as blue.

| Expts | Avg Sensitivity | Avg Precision | Avg Mag Accuracy | Avg Volumetric Similarity | Avg AUC | Average Hausdorff Distance | Average Mahanabolis Distance | ICC (#ePVS) | ICC (Volume) |
|---|---|---|---|---|---|---|---|---|---|
| T2w | 0.72 | 0.70 | 1.05 | 0.73 | 0.72 | 4.08 | 2.10 | 0.62 | 0.43 |
| T2w+FLAIR | 0.73 | 0.61 | 0.99 | 0.71 | 0.72 | 4.72 | 4.82 | 0.65 | 0.62 |
| T2w+FLAIR +T1w | 0.72 | 0.70 | 1.03 | 0.72 | 0.71 | 2.92 | 5.88 | 0.66 | 0.64 |
| T2w+FLAIR +T1w+SWI | 0.66 | 0.74 | 1.13 | 0.73 | 0.76 | 2.41 | 0.84 | 0.74 | 0.63 |
| T2w+T1w | 0.53 | 0.74 | 0.97 | 0.64 | 0.61 | 4.16 | 1.37 | 0.76 | 0.74 |
| T1w | 0.28 | 0.48 | 0.61 | 0.35 | 0.51 | 8.75 | 3.53 | 0.50 | 0.52 |
| FLAIR | 0.58 | 0.64 | 0.90 | 0.59 | 0.61 | 4.05 | 1.56 | 0.27 | 0.06 |
| T1w+FLAIR | 0.72 | 0.61 | 1.01 | 0.77 | 0.75 | 3.93 | 5.29 | 0.50 | 0.44 |

**Table S9** - Subject-wise evaluation in the occipital region. The best scores are marked as red and the second best as blue.

| Expts | Avg Sensitivity | Avg Precision | Avg Mag Accuracy | Avg Volumetric Similarity | Avg AUC | Average Hausdorff Distance | Average Mahanabolis Distance | ICC (#ePVS) | ICC (Volume) |
|---|---|---|---|---|---|---|---|---|---|
| T2w | 0.71 | 0.76 | 1.06 | 0.70 | 0.70 | 3.65 | 0.55 | 0.87 | 0.68 |
| T2w+FLAIR | 0.70 | 0.72 | 1.03 | 0.68 | 0.70 | 3.69 | 0.63 | 0.78 | 0.76 |
| T2w+FLAIR +T1w | 0.68 | 0.68 | 0.98 | 0.63 | 0.68 | 4.24 | 0.57 | 0.74 | 0.83 |
| T2w+FLAIR +T1w+SWI | 0.49 | 0.64 | 0.82 | 0.59 | 0.57 | 4.70 | 0.71 | 0.80 | 0.85 |
| T2w+T1w | 0.21 | 0.49 | 0.54 | 0.39 | 0.51 | 9.49 | 1.23 | 0.80 | 0.76 |
| T1w | 0.53 | 0.59 | 0.82 | 0.69 | 0.58 | 5.20 | 0.70 | 0.50 | 0.30 |
| FLAIR | 0.67 | 0.74 | 1.02 | 0.71 | 0.69 | 3.87 | 0.57 | 0.16 | 0.05 |
| T1w+FLAIR | 0.65 | 0.70 | 0.98 | 0.67 | 0.69 | 4.10 | 0.53 | 0.58 | 0.50 |